\documentclass[12pt,english]{article}
\usepackage{latexsym}
\usepackage{amssymb}
\usepackage{epsfig}
\usepackage{fancyhdr}
\usepackage{float}
\usepackage[T1]{fontenc}
\usepackage{graphics}
\usepackage{graphicx}
\usepackage[latin1]{inputenc}
\usepackage{multicol}
\usepackage{pslatex}
\usepackage{pifont}
\usepackage{rotating}
\usepackage{supertabular}
\usepackage{t1enc}
\usepackage{units}
\usepackage{varioref}
\usepackage{wrapfig}
\usepackage{subfigure}
\def\0{\phantom{0}}

\begin{document}

\pagenumbering{arabic}
\baselineskip25pt

\begin{center}
{\bf \large Henry's Law Constants of Methane, Nitrogen, Oxygen and Carbon Dioxide in Ethanol from 273 to 498~K: \\Prediction from Molecular Simulation}\\
\bigskip

\renewcommand{\thefootnote}{\fnsymbol{footnote}}
 Thorsten Schnabel, Jadran Vrabec\footnote{author to whom correspondence should be addressed, Tel.: +49-711/685-6107, Fax:
+49-711/685-6140, Email: vrabec@itt.uni-stuttgart.de}, Hans Hasse \\
\renewcommand{\thefootnote}{\arabic{footnote}}

Institut f\"ur Technische Thermodynamik und Thermische Verfahrenstechnik, \\
Universit\"at Stuttgart, D-70550 Stuttgart, Germany \\
\end{center}

\begin{abstract}
\baselineskip25pt
\noindent Henry's law constants of the solutes methane, nitrogen, oxygen and carbon dioxide in the solvent ethanol are predicted by molecular simulation. The molecular models for the solutes are taken from previous work. 
For the solvent ethanol, a new rigid anisotropic united atom molecular model based on Lennard-Jones and Coulombic interactions is developed. It is adjusted to experimental pure component saturated liquid density and vapor pressure data. Henry's law constants are calculated by evaluating the infinite dilution residual chemical potentials of the solutes from 273 to 498~K with Widom's test particle insertion. The prediction of Henry's Law constants without the use of binary experimental data on the basis of the Lorentz-Berthelot combining rule agree well with experimental data, deviations are 20~\%, except for carbon dioxide for which deviations of 70~\% are reached. Quantitative agreement is achieved by using the modified Lorentz-Berthelot combining rule which is adjusted to one experimental mixture data point.
\end{abstract}

\bigskip

\textbf{Keywords:}  Henry's law constant, molecular simulation 

\bigskip

\clearpage

\section{Introduction}

The goal of the 2nd Industrial Fluid Properties Simulation Challenge 2004 organized by the American Institute of Chemical Engineers (AIChE) is ''to assess the current abilities and inabilities in the prediction of  physical properties applying force fields and molecular techniques'' \cite{www_sc}. One problem proposed in this Challenge is the prediction of Henry's law constants for the solutes methane, nitrogen, oxygen and carbon dioxide in the solvent ethanol at the temperatures 323 and 373~K. This problem is tackled here using multi-center Lennard-Jones, polar and electrostatic interaction potentials together with molecular simulation.

For modeling the solutes, the symmetric two-center Lennard-Jones plus pointpolarity potential is used. The parameters of this model were adjusted for 80 pure fluids to experimental saturated liquid density and vapor pressure data in previous work of our group \cite{Vrabec2001,Vrabec2004}.

Many molecular models for ethanol are available in the literature, some of them consider all atoms explicitly, some of them use the united atom approach for methyl and methylene groups. Most of the existing models account for internal degrees of freedom, but only few explicitly for polarizability effects. An overview of ethanol models from the literature is given in section \ref{solvent1}.

Among the available molecular models for ethanol, the probably most appropriate for predicting vapor-liquid coexistence properties is the transferable potential for phase equilibria-united atom (TraPPE-UA) of Chen et al.~\cite{Chen2001}. That model accounts for internal degrees of freedom. In the present work, a new simple molecular model for ethanol was developed with the aim to yield at least as accurate results but with distinctly less computational and programming effort. The new rigid  effective ethanol model is of the united-atom type and based on Lennard-Jones and Coulombic interactions. The united atom approach and the neglect of internal degrees of freedom saves computation time which is desirable for applications of force field methods in industrially relevant problems.

An infinitely diluted solute in a solvent only experiences solute-solvent interactions. The results for Henry's law constants from molecular simulations therefore depend on the model used for describing the unlike solute-solvent interaction. Consequently, predictions of the Henry's law constant from pure component data alone are a hard test for every molecular model. The unlike interactions are usually determined from the like interactions which are known from pure component models by combining rules like that of Lorentz-Berthelot. These combining rules usually only have a weak theoretical basis so that it is an open question how useful they are for predictions of Henry's law constants from pure component data alone. On that background it makes sense to also study modified combining rules which allow an adjustment of the model for the unlike interaction to experimental mixture data. Experience shows that if such an adjustment is carried out even for only one single mixture data point, excellent predictions of mixture properties can be achieved over a wide range of states \cite{Vrabec1995}-\cite{Stoll2003}.

Henry's law constants were predicted by applying the molecular models which were parameterized exclusively to pure substance vapor-liquid equilibria and by using the Lorentz-Berthelot combining rule for the unlike Lennard-Jones interactions. Also some other combining rules are briefly discussed. Additionally, the modified Lorentz-Berthelot combining rule with one state independent binary adjustable parameter is investigated.

\section{Molecular Models}
\subsection{Ethanol models from the literature \label{solvent1}}

In this section, an overview of molecular ethanol models from the literature based on the Lennard-Jones potential and point charges is given. The models are assessed regarding their description of vapor-liquid equilibria where such information is available in the literature. In 1981, Jorgensen \cite{Jorgensen1981} presented a rigid ethanol model based on transferable intermolecular potential functions (TIPS). TIPS consider the methyl and methylene groups as single Lennard-Jones sites, centered on the carbon nuclei and a single Lennard-Jones site centered on the oxygen nucleus accounting for the dispersion and repulsion of the hydroxyl group. Point charges are located on the hydroxyl hydrogen and the bonded oxygen nucleus. The charge located on the methylene group is chosen to achieve overall neutrality of the molecule. That model was developed to yield reasonable structural and energetic results. Subsequently, Jorgensen investigated the effect of internal rotation in the ethanol molecule about the carbon-oxygen bond \cite{Jorgensen1981b}, regarding the structure and hydrogen bonding and compared the results to X-ray data for the solid and to ab initio molecular orbital calculations.

Improvements of the Lennard-Jones parameters of the united alcyl \cite{Jorgensen1984} and hydroxyl groups, the point charges, the internal rotational potential and the geometry, were achieved by fitting directly to experimental thermodynamic and structural data, as well as quantum chemical molecular mechanics calculations. This yielded the united-atom optimized potential model for liquid simulations (OPLS-UA) \cite{Jorgensen1986}. Vapor-liquid equilibria with the OPLS-UA were calculated by van Leeuwen \cite{Leeuwen1996}. A comparision of the these simulation results with experimental data shows a mean unsigned error \footnote[1]{mean unsigned error: $\frac{1}{M} \sum\limits_{i=1}^{M} \frac{|z_i^{\rm sim}- z_i^{\rm exp}|}{z_i^{\rm exp}} $, where $M$ is the number of simulated data, $z_i^{\rm sim}$ is the simulated and $z_i^{\rm exp}$ the experimental quantity of interest.} of about 30~\% in vapor pressure, 6~\% in saturated liquid density and 7~\% in heat of vaporization. Stubbs et al. \cite{Stubbs2001} suggested a modification of the Lennard-Jones parameters of the OPLS-UA methyl group. Petravic and Delhommelle~\cite{Petravic2003} termed this modified model S2 and compared it to the OPLS-UA regarding liquid density, hydrogen bonding and diffusion. From that study, it was concluded that the S2 model improves the prediction of liquid density. However, the S2 model overestimates the diffusion coefficient by approximately 15~\%, whereas the OPLS-UA gives better results. Also the influence of the internal degrees of freedom on transport quantities has been studied by Petravic and Delhommelle. They showed that the density is affected only slightly by the internal degrees of freedom, in contrast to diffusion coefficients.    

More detailed potential models for ethanol consider all atoms explicitly. Müller-Plathe \cite{Mueller-Plathe1996} designed such an all atom model including internal rotation about the carbon-oxygen bond to give reasonable bulk properties in mixtures with water. Jorgensen et al.~\cite{Jorgensen1996} published an OPLS all atom potential model (OPLS-AA) considering additionally stretching, bending and internal rotation about the carbon-carbon and carbon-oxygen bond. A similar molecular model which uses the same approach as OPLS-AA but with a different parameter set is the AMBER94 force field \cite{Cornell1995,Cornell1996}. Panagiotopoulos \cite{Panagiotopoulos2000} concluded that the OPLS models are not appropiate predicting vapor-liquid coexistence since they were developed for energetic and structural properties at ambient conditions. 

Optimized models for vapor-liquid equilibria of linear and branched alkanes were developed by Martin and Siepmann \cite{Martin1998,Martin1999}, i.e.~transferable potentials for phase equilibria-united atom (TraPPE-UA). The TraPPE-UA was extended to alcohols \cite{Chen2001} which gives to our knowledge the most accurate existing ethanol model for vapor-liquid equilibria.  It was found that this model yields vapor pressures, saturated liquid densities and heats of vaporization with mean unsigned errors of about 8, 1 and 1~\%, respectively.

Since polarazible molecular models are computationally very expensive, only few ethanol models of that type are available. The polaraziable intermolecular potential function (PIPF) proposed by Gao et al. \cite{Gao1995} is based on the reparamerization of the OPLS-AA plus isotropic point polarization on each atom. That model was developed to yield better results for heat of vaporization and density. Another polarizable model is that from González et al. \cite{Gonzalez1999}. It is based on the OPLS-UA and consists of only one isotropic point polarizability on the (pseudo) oxygen atom. Delhommelle et al.~\cite{Delhommelle2000a} showed for pentane and hydrogen sulfide that polarizable models can give better predictions of mixture vapor-liquid equilibria.

\subsection{New ethanol model\label{solvent2}}

A new ethanol model was developed with the aim to give accurate vapor-liquid equilibria with low computational and programming effort. It neglects the internal degrees of freedom and uses nuclei off-center Lennard-Jones united atoms for the methyl, methylene and hydroxyl group, accounting for repulsion and dispersion. Point charges are located on the methylene and hydroxyl Lennard-Jones centers, as well as on the nucleus position of the hydroxyl hydrogen. The Coulombic interactions account for both polarity and hydrogen bonding. 

The molecular interaction and geometry of the resulting model is described in the following. The potential energy $u_{ij}$ between two ethanol molecules $i$ and $j$ is given by

\begin{equation}
	u_{ij}\left( r_{ijab}\right) =\sum_{a=1}^{4}\sum_{b=1}^{4}4\epsilon_{ab}\left[ \left( \frac{\sigma_{ab}}{r_{ijab}}\right)^{12}-\left( \frac{\sigma_{ab}}{r_{ijab}}\right)^6\right] + \frac{q_{ia}q_{jb}}{4\pi\varepsilon_{0}r_{ijab}}\, , \label{equ_etha}
\end {equation}

\noindent where $a$ is the site index of molecule $i$ and $b$ the site index of molecule $j$, respectively. The site-site distance between molecules $i$ and $j$ is denoted $r_{ijab}$. $ \sigma_{ab} $, $\epsilon_{ab}$ are the Lennard-Jones size and energy parameters,  $ q_{ia}$ and $q_{jb}$ are the point charges located at the sites $ a $ and $b$ on the molecules $i$ and $j$, respectively.  Finally, $\varepsilon_{0}$ is the permittivity of vacuum. The interaction between unlike Lennard-Jones sites of two ethanol molecules is defined by the Lorentz-Berthelot combining rules

\begin{eqnarray}
	\sigma_{ab}   & = & \frac{\sigma_{aa}+\sigma_{bb}}{2}\,,\\
	\epsilon_{ab} & = & \sqrt{\epsilon_{aa}\epsilon_{bb}}\,. \label{LBCR}
\end{eqnarray}

\begin{figure}[ht]
\epsfig{file=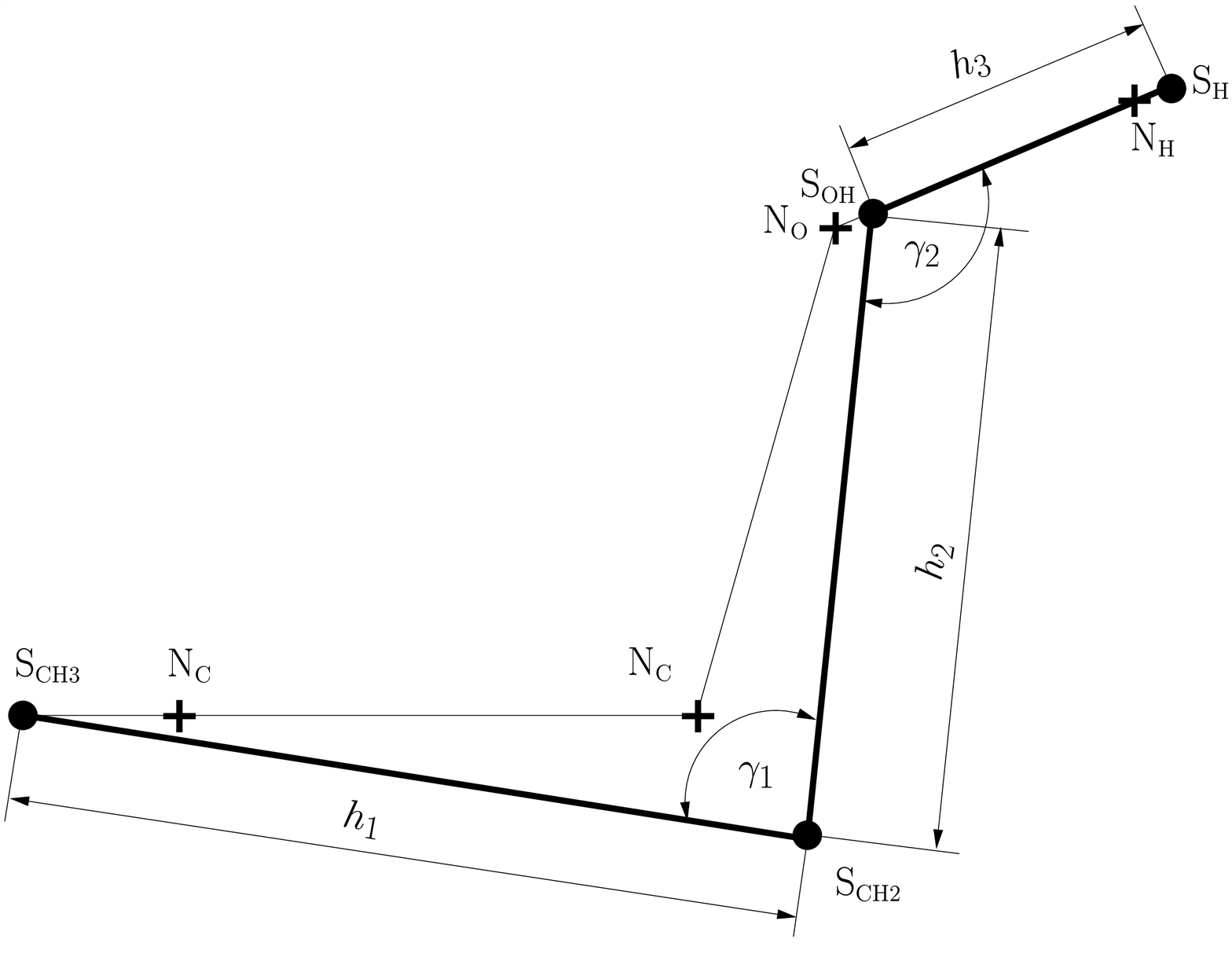,scale=0.5}
\bigskip
\bigskip
\bigskip
\bigskip
\caption[Geometry of the new ethanol model: $\rm{S}_i$ indicates the site $i$ and $\rm{N}_j$ the nucleus position of atom $j$ obtained from quantum chemistry calculations.]{\label{geo_etha}}
\end{figure}

\noindent The geometry of the present ethanol model and the potential parameters are given in Figure \ref{geo_etha} and Table \ref{tab_etha_pot}.

In the following, a few remarks on the model development are given. Jorgensen's investigations on the potential of the internal rotation about the carbon-oxygen bond in an ethanol monomer \cite{Jorgensen1981b} yield three energetic extremal values in the corresponding configurations: the absolute minimum is reached in the trans configuration (dihedral angle $\phi=180°$), the 0.64 kcal/mol higher local minimum in the gauche configuration ($\phi\approx 60°$) and the 2.06 kcal/mol (above trans) higher absolute maximium in the cis configuration ($\phi=0°$). For the present rigid ethanol model, the energetically most likely trans configuration has been chosen. The nuclei positions of all ethanol atoms were computed using the quantum chemistry software package GAMESS (US) \cite{Schmidt1993}. The basis set 6-31G and the Hartree-Fock method were applied for geometry optimization. The distances between the carbon-carbon, carbon-oxygen and oxygen-hydroxyl hydrogen nuclei positions obtained from quantum chemistry calculations are 1.51342, 1.47986 and 0.95053 \r{A}, respectively. The CCO and COH angles are 110.760 and $117.280°$. Figure \ref{geo_etha} includes also these nuclei positions.

Starting from these positions, for the methyl and methylene group the AUA4 parameters of Ungerer et al.~\cite{Ungerer2000} were applied. The AUA4 parameters were optimized by Ungerer et al.~for vapor-liquid equilibria of linear alkanes. Following the approach of Ungerer et al.~\cite{Ungerer2000}, a small offset of the Lennard-Jones hydroxyl center (0.1010~\r{A}) in direction to the hydroxyl hydrogen nucleus was allowed for optimization. The five parameters of the hydroxyl group were fitted to yield optimal saturated liquid densities and vapor pressures. The parameters are the two point charges, the Lennard-Jones size and energy parameters as well as the offset of the hydroxyl Lennard-Jones center. These parameters were chosen since they model the strongly interacting hydrogen bond and since the methyl and methylene parameters were already optimized by Ungerer et al.~\cite{Ungerer2000}. The point charge on the methylene center was set in such a manner to yield overall neutrality of the molecular model.

The proposed set of charges in the present work yields a resulting dipole moment of 2.47 D. Quantum chemistry calculations (M{\o}ller-Plesset level 2 with basis set aug-cc-pVTZ) in a cavity (COSMO with a dielectric constant $\epsilon_{\rm r}=25$) and in the vacuum yield dipole moments of 1.99 and 2.10 D, respectively. Note that the higher dipole moment of the molecular model than those from quantum chemistry calculations is not unusual for modeling hydrogen bonding molecules with point charges.

Pure substance vapor-liquid equilibria for the optimization of the ethanol model were calculated using the $NpT$+test particle method proposed by M\"{o}ller and Fischer \cite{Moeller1990}-\cite{Moeller1994}. To determine the residual chemical potential in the liquid phase with high accuracy, the Monte Carlo based gradual insertion method \cite{Shevkunov1988}-\cite{Vrabec2002a} was used. For the residual chemical potential in the vapor phase, standard Monte Carlo simulations with Widom's test particle insertion \cite{Widom1963} proved to be sufficient. Simulation details are given in the Appendix.  

\begin{figure}[ht]
\epsfig{file=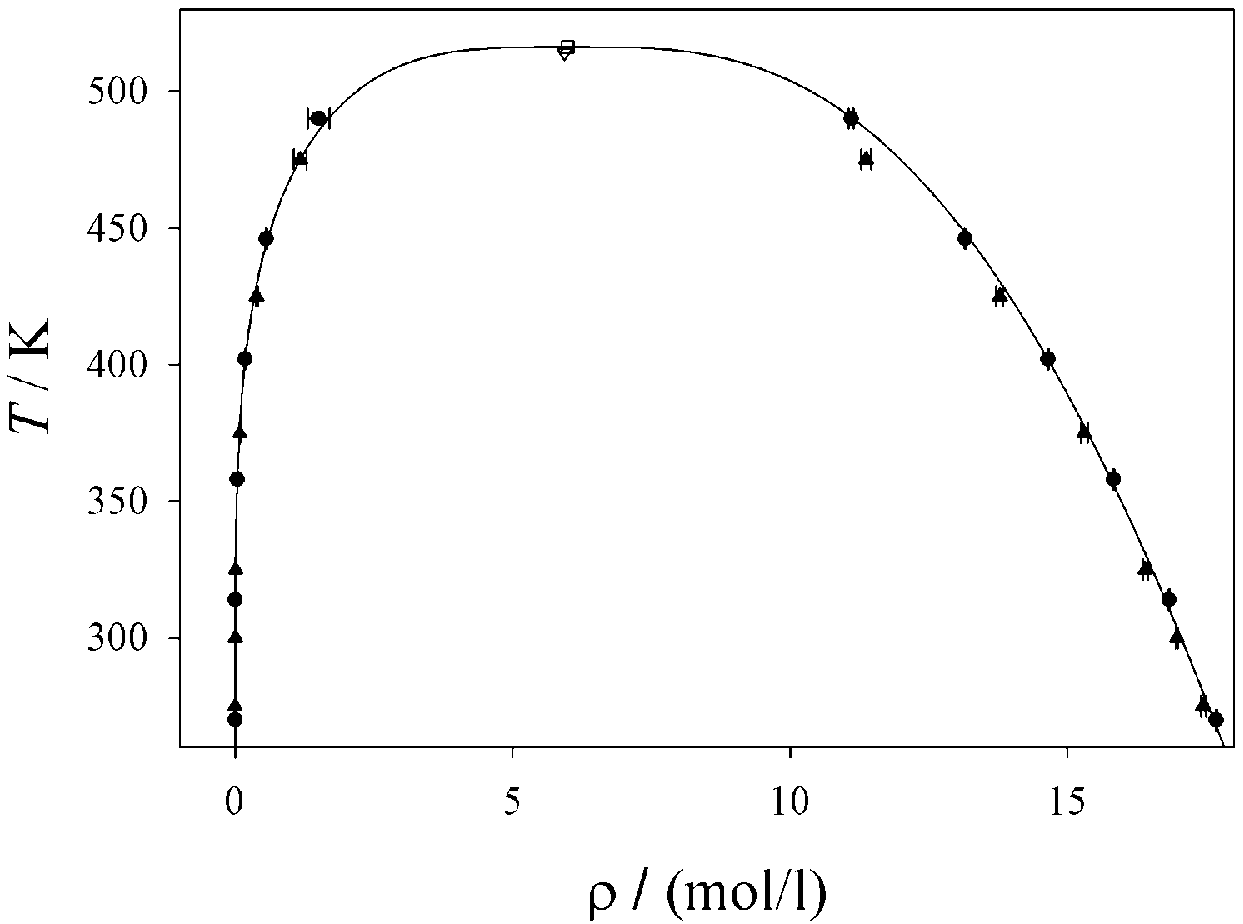,scale=0.5}
\bigskip
\bigskip
\bigskip
\bigskip
\caption[Saturated densities of ethanol: {\Large $\bullet$}  present simulation, $\triangledown$ critical point derived from simulated data, $\blacktriangle$ TraPPE-UA simulation \cite{Chen2001}, {---}~experimental data \cite{Daubert1984}, $\square$ experimental critical point \cite{Daubert1984}.]{\label{fig_TRho_etha}}
\end{figure}

\begin{figure}[ht]
\epsfig{file=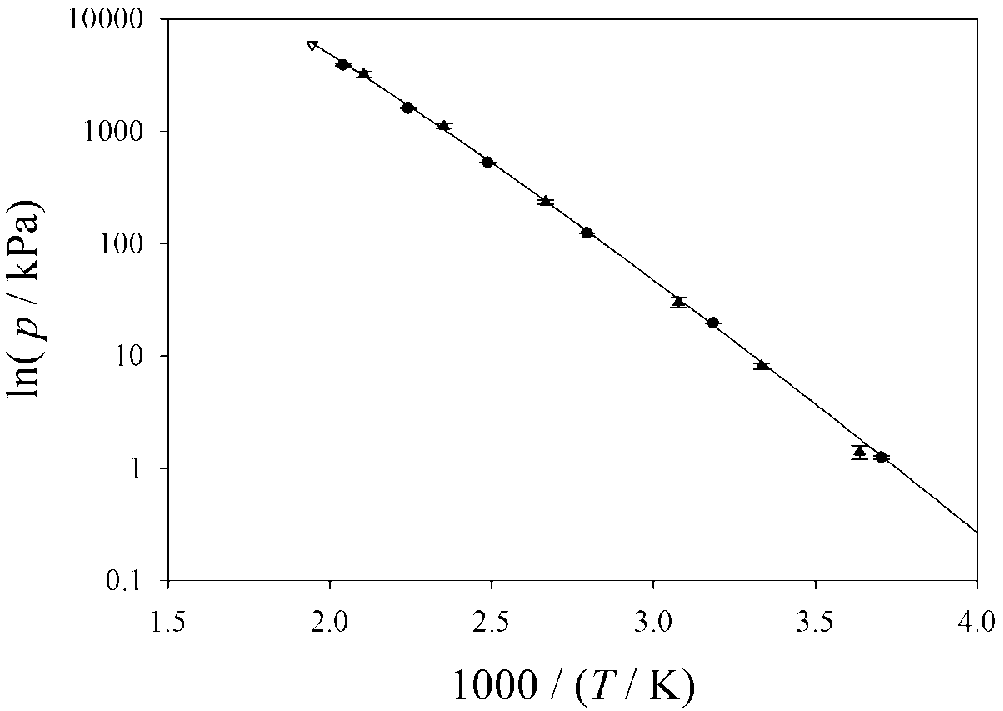,scale=0.5}
\bigskip
\bigskip
\bigskip
\bigskip
\caption[Vapor pressure of ethanol: {\Large $\bullet$}  present simulation, $\triangledown$ critical point derived from simulated data, $\blacktriangle$ TraPPE-UA simulation \cite{Chen2001}, {---}~experimental data \cite{Daubert1984}.]{\label{fig_pT_etha}}
\end{figure}

\begin{figure}[ht]
\epsfig{file=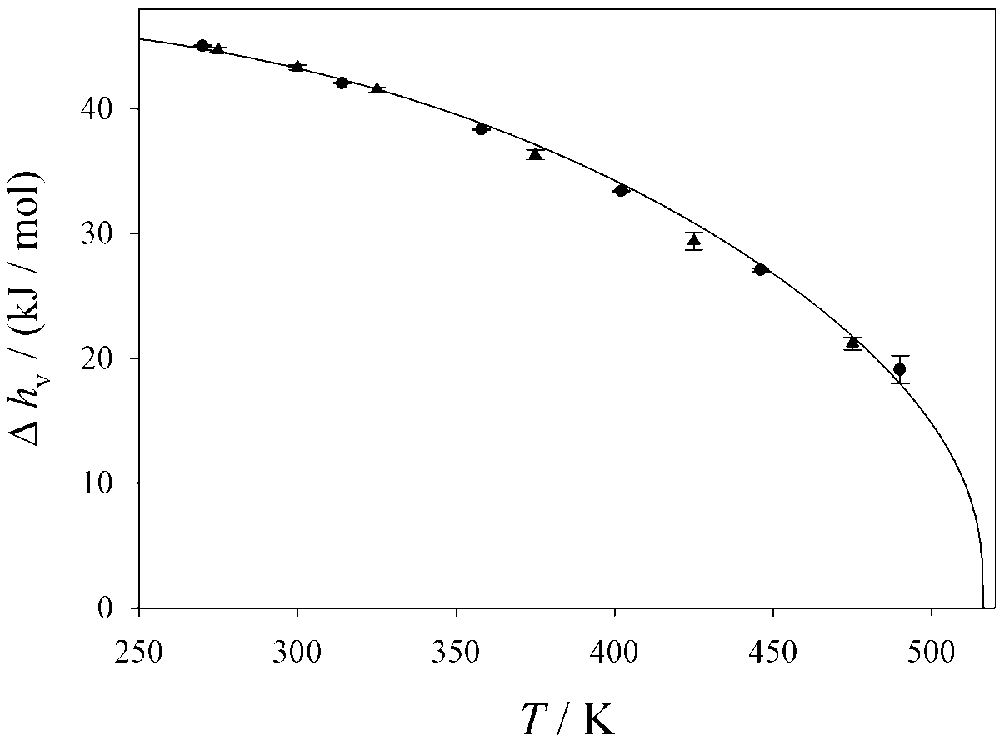,scale=0.5}
\bigskip
\bigskip
\bigskip
\bigskip
\caption[Heat of vaporization of ethanol: {\Large $\bullet$}  present simulation, $\blacktriangle$ TraPPE-UA simulation \cite{Chen2001}, {---}~experimental data \cite{Daubert1984}.]{\label{fig_dhv_etha}}
\end{figure}

Results for vapor-liquid equilibria obtained with the new ethanol model together with experimental data \cite{Daubert1984} are given in Table \ref{tab_eqr_etha}. Figures \ref{fig_TRho_etha}, \ref{fig_pT_etha} and \ref{fig_dhv_etha} depict the results including the TraPPE-UA simulation data taken from \cite{Chen2001}. The agreement between the new model and the experimental data is excellent, even better than that for the more complex and already very good TraPPE-UA ethanol model. The simulaltion results of the present ethanol model yield mean unsigned errors compared to experimental data \cite{Daubert1984} in vapor pressure, saturated liquid density and heat of vaporization of 3.7, 0.3, and 0.9~\%, respectively, in the temperature range 270 to 490~K, which is about 55 to 95~\% of the critical temperature. Following the procedure suggested by Lotfi et al.~\cite{Lotfi1992}, the critical temperature, density and pressure were determined. The results compare favorably to experimental data (numbers in parenthesis): $T_c=$514.12 (516.25)~K, $\rho_c=$5.94 (5.99)~mol/l and $p_c=$5.89 (6.38)~MPa. The TraPPE-UA molecular model for ethanol yields mean unsigned errors in vapor pressure, saturated liquid density and heat of vaporization of 8.4, 1.4, and 1.6~\%, respectively. Critical temperature and density are quoted with $T_c=$514 K and $\rho_c=$6.1 ~mol/l \cite{Chen2001}.

\subsection{Solutes \label{solutes}}

Numerous Lennard-Jones based potential models for the solutes studied in the present work are available from the literature, 
e.g.~for methane 
\cite{Vrabec1995,Jorgensen1984,Martin1998},\cite{Fischer1984}-\cite{McDonald1972}, 
for nitrogen 
\cite{Fischer1984}-\cite{Rivera2000},
for oxygen 
\cite{Hirschfelder1954,McDonald1972,Bohn1985,Jelinek1972,Bohn1986} 
and for carbon dioxide 
\cite{Moeller1994,Fischer1984,Hirschfelder1954,MacRury1976},\cite{Gibbons1974}-\cite{Suzuki1971}. 

Here, models of the symmetric two-center Lennard-Jones plus pointquadrupole (2CLJQ) type \cite{Vrabec2001} were choosen, since that model class is able to describe the intermolecular interactions of all four solute  molecules consistently. Furthermore, it has succesfully been applied to mixtures \cite{Vrabec2004,Stoll2003}. The potential parameters were adjusted in previous work of our group \cite{Vrabec2001,Vrabec2004} exclusively to experimental pure component vapor-liquid equilibria, i.e.~bubble density and vapor pressure. The complete parameter set of the solute models is listed in Table \ref{tab_2cljq}. A detailed definition of the symmetric elongated quadrupole with momentum $Q$ is given in \cite{Vrabec2004}. The mean unsigned error of the vapor pressure, saturated liquid density and heat of vaporization is smaller than 1, 3 and 3~\%, respecitvely, for all four solute models practically over the whole vapor-liquid equilibrium temperature range.

\subsection{Mixtures \label{mix}}

In molecular simulations of binary mixtures with pairwise additive potentials, three different interactions are present: two between molecules of the same component which are fully defined by the pure substance models and the unlike interaction between molecules of different kind. The unlike polar interactions were determined in a straightforward manner without using binary parameters. To define the unlike Lennard-Jones interactions between solvent and solutes methane, nitrogen, oxygen and carbon dioxide, the modified Lorentz-Berthelot combining rule was used

\begin{eqnarray}
	\sigma_{ab}   & = & \frac{\sigma_{aa}+\sigma_{bb}}{2} \label{mLBcr1}\, ,\\
	\epsilon_{ab} & = & \xi \cdot \sqrt{\epsilon_{aa}\epsilon_{bb}}  \, ,\label{mLBcr2}
\end{eqnarray}

\noindent where $\xi$ is the binary interaction parameter which accounts for mixture effects. 

In the ``predictive mode'', the Henry's law constants were computed setting $\xi=1$, cf. Equation (\ref{mLBcr2}). 

Simulations of many binary mixtures \cite{Vrabec2004,Stoll2003} have shown that the Lorentz-Berthelot combining rule ($\xi$=1) is too crude to predict the unlike Lennard-Jones energy parameter in an accurate way. Therefore, in the ``adjusted mode'', $\xi$ was fitted to one experimental Henry's law constant. Compared to using polarizable models, the adjustment of the state independent parameter $\xi$ is a simple procedure \cite{Stoll2003}. The results for the predictive and the adjusted mode calculating Henry's law constants are discussed in Section \ref{rnd}.  

Other combining rules were suggested e.g.~by Kohler \cite{Kohler1957}, Hudson and McCoubrey \cite{Hudson1960}, Fender and Halsey \cite{Fender1962}, Hiza et al.~\cite{Hiza1969}-\cite{Hiza1978}, Smith and Kong \cite{Smith1972,Kong1973}, Waldman and Hagler \cite{Waldman1993}, Al-Matar and Rockstraw \cite{Al-Matar2004}. They were mainly developed on the basis of noble gas data. The unlike Lennard-Jones energy parameters of binary noble gas mixtures determined from the above mentioned combining rules are mostly smaller than those given by the geometric mean of the Lorentz-Berthelot combining rule. But as shown in Section \ref{rnd}, the unlike solvent-solute interaction determined from the Lorentz-Berthelot combining rule has to be increased for three out of the four studied solutes ($\xi>1$) to yield quantitative results. Therefore, the alternative combining rules will not give better predictions and were not studied further.

\section{Henry's law constants \label{henry}}

Several approaches have been proposed in the literature \cite{Murad2000,Sadus1997} to obtain Henry's law constants on the basis of molecular models. Henry's law constants are related to the residual chemical potential of the solute $i$ at infinite dilution ${\mu_i}^{\infty}$ \cite{Shing1988}

\begin{eqnarray}
H_i={\rho}k_{\rm B}T \exp{({\mu_i}^{\infty}/(k_{\rm B}T))}\, ,\label{hsim}
\end{eqnarray}

\noindent where $k_{\rm B}$ is the Boltzmann constant, $T$ the temperature, $\rho$ the density of the solvent.

In order to evaluate ${\mu_i}^{\infty}$, molecular dynamics simulations applying Widom's test particle method \cite{Widom1963} are sufficient here. This is due to the fact, that the solute molecules are all smaller than ethanol molecules and so acceptable statistics can be achieved. Therefore, test particles representing the solute $i$ were inserted after each time step at random positions into the liquid solvent and the potential energy between the solute test particle and all solvent molecules ${\psi}_i$ was calculated within the cut-off radius

\begin{eqnarray}
{\mu_i}^{\infty}=-k_{\rm B}T\ln{\langle}V\exp({-\psi_i/(k_{\rm
B}T)){\rangle}}/{\langle}V{\rangle},\label{mu}
\end{eqnarray}

\noindent where $V$ is the volume and the brackets represent the $NpT$ ensemble average.

The residual chemical potential at infinite dilution and hence, the Henry's law constant is directly attributed to the unlike solvent-solute interaction and indirectly to the solvent-solvent interaction which yields the configurations of the solvent molecules. In these configurations, the solvent test particles are inserted. The mole fraction of the solute in the solvent is exactly zero, as required for infinite dilution, since the test particles are instantly removed after the potential energy calculation. Simulations were performed at specified temperature and the according vapor pressure of pure ethanol.

\section{Results\label{rnd}}

Henry's law constants of the four solutes methan, nitrogen, oxygen and carbon dioxide in ethanol were determined from simulation in the predictive mode at temperatures between 273 to 498~K with an increment of 25~K. These simulation results are given in Table \ref{tab_pred}.

\begin{figure}[ht]
\epsfig{file=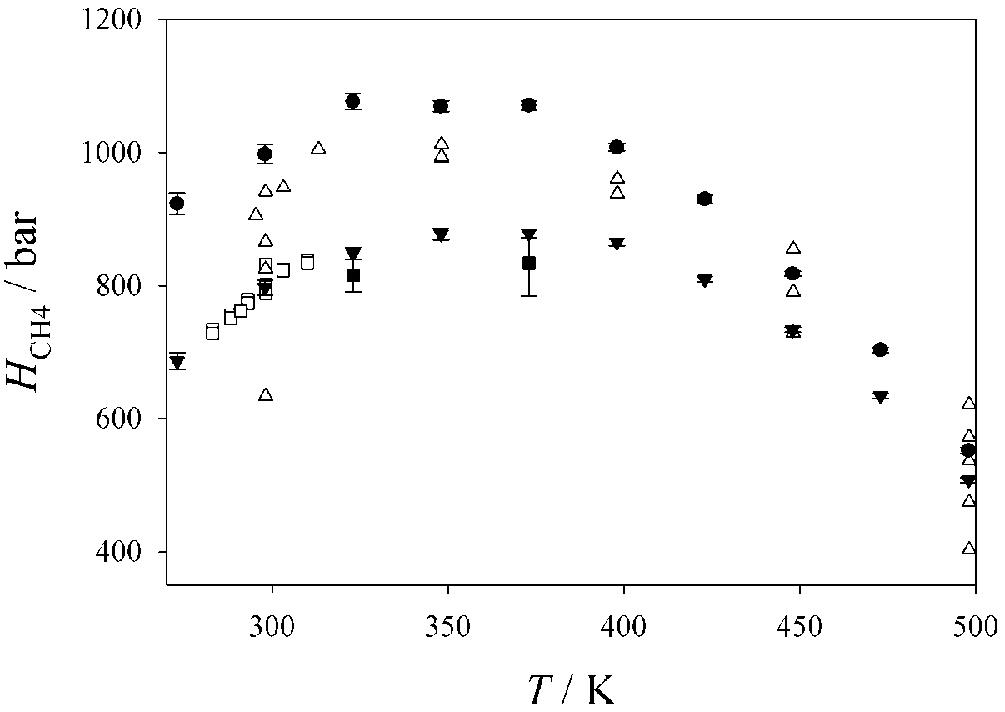,scale=0.5}
\bigskip
\bigskip
\bigskip
\bigskip
\caption[Henry's law constants of methane in ethanol: $\square$ confirmed experimental data \cite{Yaacobi1974},\cite{Boyer1960}-\cite{Lannung1960}, $\vartriangle$ other experimental data \cite{McDaniel1911}-\cite{Brunner1990},  {\Large $\bullet$} predictive ($\xi=1$) simulation data,  $\blacktriangledown$ adjusted ($\xi=1.0403$) simulation data, $\blacksquare$ recently published recommended experimental data \cite{Olson2004}.]{\label{fig_h_ch4}}
\end{figure}

\begin{figure}[ht]
\epsfig{file=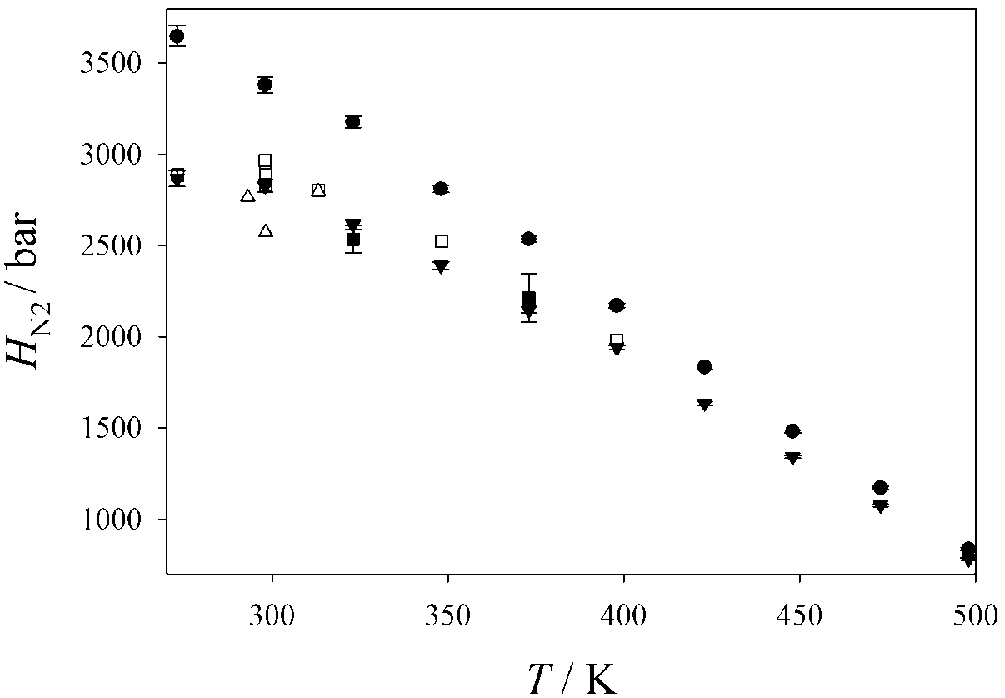,scale=0.5}
\bigskip
\bigskip
\bigskip
\bigskip
\caption[Henry's law constants of nitrogen in ethanol: $\square$ confirmed experimental data \cite{Nitta1978,Boyer1960,Bo1993,Wilken1999,Katayama1976}, $\vartriangle$ other experimental data \cite{Makranczy1979,Tokunaga1975a}, {\Large $\bullet$} predictive ($\xi=1$) simulation data,  $\blacktriangledown$ adjusted ($\xi=1.0449$) simulation data, $\blacksquare$ recently published recommended experimental data \cite{Olson2004}.]{\label{fig_h_n2}}
\end{figure}

\begin{figure}[ht]
\epsfig{file=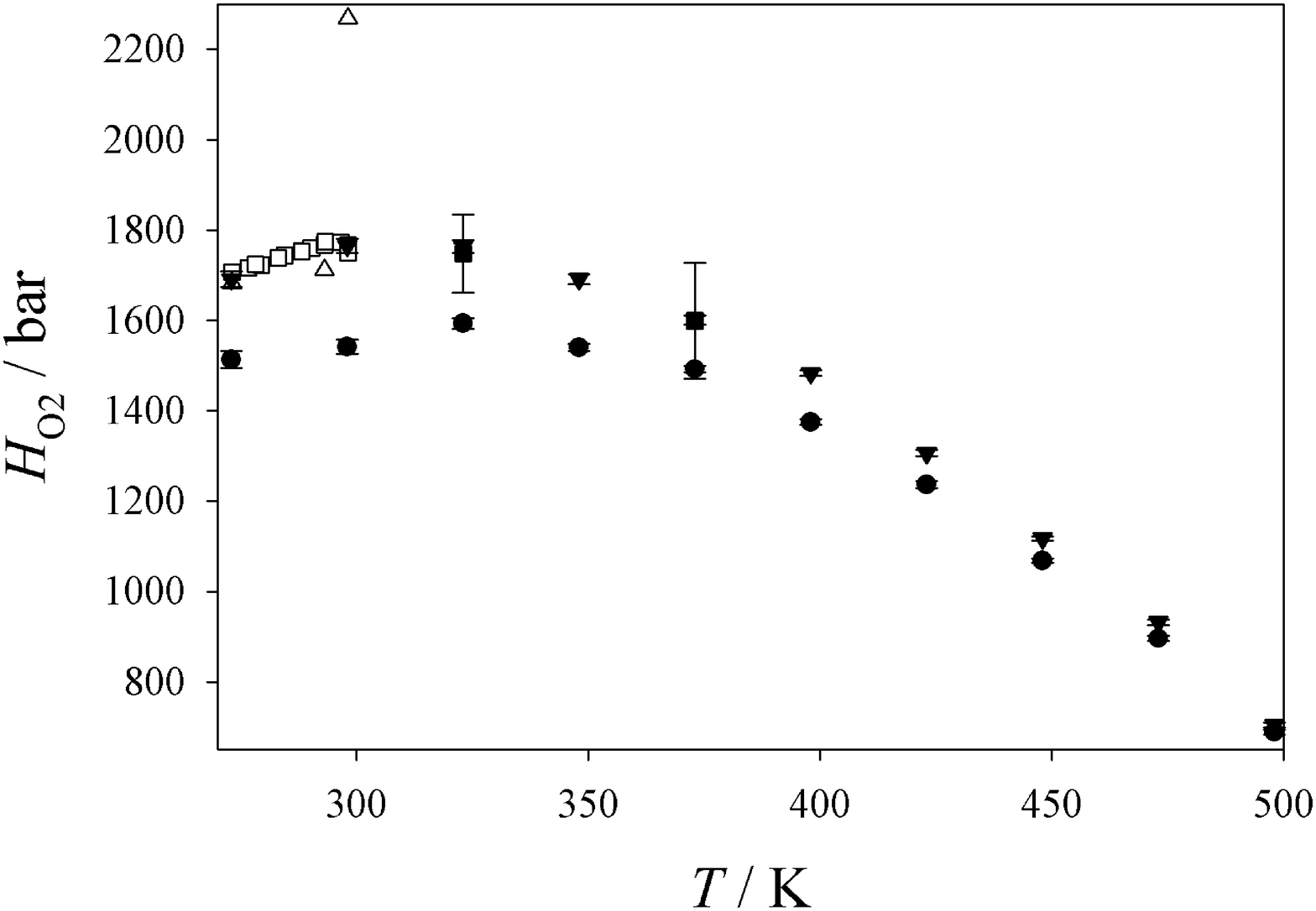,scale=0.5}
\bigskip
\bigskip
\bigskip
\bigskip
\caption[Henry's law constants of oxygen in ethanol: $\square$ confirmed experimental data \cite{Finlayson1923,Bo1993,Timofejew1890,Luehring1989}, $\vartriangle$ other experimental data \cite{Makranczy1979,Tokunaga1975a}, {\Large $\bullet$} predictive ($\xi=1$) simulation data,  $\blacktriangledown$ adjusted  ($\xi=0.9802$) simulation data, $\blacksquare$ recently published recommended experimental data \cite{Olson2004}.]{\label{fig_h_o2}}
\end{figure}

\begin{figure}[ht]
\epsfig{file=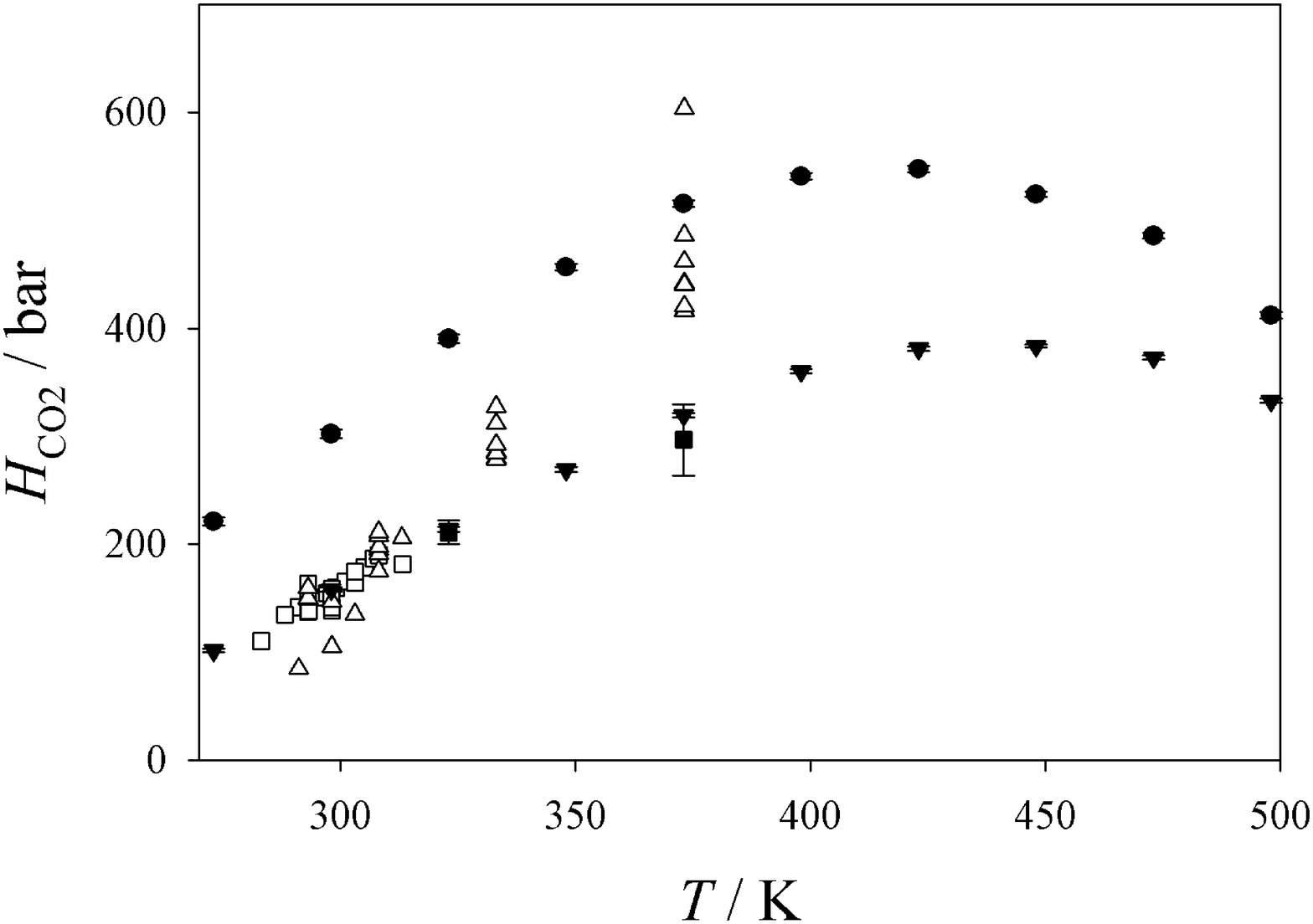,scale=0.5}
\bigskip
\bigskip
\bigskip
\bigskip
\caption[Henry's law constants of carbon dioxide in ethanol: $\square$ confirmed experimental data \cite{Postigo1987},\cite{Luehring1989}-\cite{Kierzkowska-Pawlak2002}, $\vartriangle$ other experimental data \cite{Makranczy1979,Sander1912,Chang1998},  {\Large $\bullet$} predictive ($\xi=1$) simulation data,  $\blacktriangledown$ adjusted ($\xi=1.0790$) simulation data, $\blacksquare$ recently published recommended experimental data \cite{Olson2004}.]{\label{fig_h_co2}}
\end{figure}

For assessing the simulation results reliable experimental data are necessary. For each of the four studied systems, the available experimental data were grouped into two categories: data with low scattering that is confirmed by data sets of other authors on one hand and the rest of the data on the other hand. Data of the first category are called ``confirmed data'' here for brevity. Experimental data and simulation results are depicted in Figures \ref{fig_h_ch4} to \ref{fig_h_co2}. The numerical values for the simulation results both predictive and adjusted mode are given in Tables \ref{tab_pred} and  \ref{tab_desc}.

The examination of the experimental data at 300~K shows, that the Henry's law constants differ from about 150 (carbon dioxide) up to 3000~bar (nitrogen), i.e.~more than one order of magnitude. Compared to this, the prediction of the gas solubility from pure component data alone are good, especially when taking into account that this property is dominated by the unlike interaction in mixtures. In the following, first the results of the predictive mode are discussed for all four solutes. After this follows the discussion for the results of the adjusted mode.

For methane, cf.~Figure \ref{fig_h_ch4}, the predictions agree well with the experimental data. Deviations are about 20~\%. It is remarkable that the temperature dependence of Henry's law constant of methane in ethanol, which shows a maximum at about 350~K, is predicted correctly. Since the slope of the Henry's law constant as a function of temperature $H(T)$ is related to the heat of solution, this means that also that caloric property is predicted well. Also for nitrogen, the agreement between the predictions and the experimental data is typically within 20~\%, cf.~Figure \ref{fig_h_n2}. Even better results are achieved for oxygen, cf.~Figure \ref{fig_h_o2}, where deviations are mostly within 10~\%. Only for carbon dioxide, the predicted results are worse, cf.~Figure \ref{fig_h_co2}. The small Henry's law constants of carbon dioxide in ethanol are predicted too high by up to about 70~\%. But even for that system, the predictions from molecular simulation lie within the range of the largely scattering experimental data. In all cases the predictions from the simulation for the slope of $H(T)$ are excellent. Considering the simplicity and empirical nature of the Lorentz-Berthelot combining rule used here, this agreement of the predictions of the Henry's law constants with the experimental data is astonishingly good. 

As explained above, combining rules have only a weak theoretical basis and work best when used in a form that allows an adjustment to experimental mixture data. For that adjustment one single reliable mixture data point is often sufficient, even when a wide range of states of the mixture needs to be modelled. All this can be demonstrated very well also for the case study of the present work. For all four solutes, reliable experimental data on the Henry's law constants are available at around room temperature. Data at other temperatures are usually questionable, except perhaps for nitrogen. Therefore, in the present work for all four solutes, the binary interaction parameter $\xi$ of the modified Lorentz-Berthelot combining rule, cf.~ Equation (\ref{mLBcr2}), was adjusted to experimental data of the Henry's law constants at 298~K (methane \cite{Yaacobi1974}, nitrogen \cite{Nitta1978}, oxygen \cite{Finlayson1923} and carbon dioxide \cite{Postigo1987}). The adjusted binary interaction parameters are $\xi=1.0403$ (methane), 1.0449 (nitrogen), 0.9820 (oxygene) and 1.0790 (carbon dioxide) and were used for predictions of the Henry's law constants at other temperatures. The results given in Table \ref{tab_desc} and shown in Figures \ref{fig_h_ch4} to \ref{fig_h_co2} are what we hold to be presently the most reliable and available information on $H(T)$ of the studied solutes in ethanol. The predictions are confirmed in the only case were the experimental data base allows such a test, i.e.~for nitrogen (Figure \ref{fig_h_n2}).

Recently recommended experimental data were published on the Internet \cite{Olson2004}. In Figures \ref{fig_h_ch4} to \ref{fig_h_co2}, these data were added together with their uncertainties which are typically about 5~\%. All simulation results of the descriptive mode agree with the recommended experimental results within the uncertainties. Hence, the estimated deviations of the Henry's law constants determined in the predictive mode are validated by these recommended data.

\begin{figure}[ht]
\epsfig{file=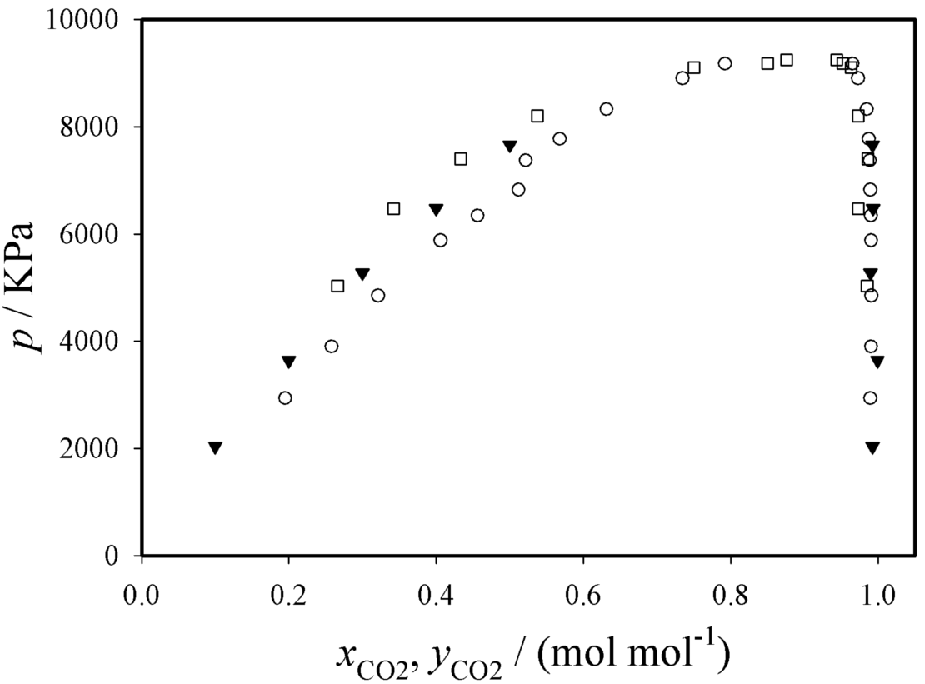,scale=0.5}
\bigskip
\bigskip
\bigskip
\bigskip
\caption[Vapor-liquid equilibria of the mixture carbon dioxide and ethanol at 323~K: $\bigcirc$ \cite{Yoa1989}, $\square$ \cite{Lim1994} experimental data, $\blacktriangledown$ adjusted ($\xi=1.0790$) simulation data.]{\label{fig_pxy}}
\end{figure}

Additionally to the calculation of the Henry's law constants, the vapor-liquid equilibria of the mixture carbon dioxide and ethanol at 323~K were simulated with the Grand Equilibrium method \cite{Vrabec2002} using the adjusted binary interaction parameter ($\xi=1.0790$). The simulation results and the experimental data are given in Figure \ref{fig_pxy}.

The excellent agreement of the decriptive mode results for both the Henry's law constants and the vapor-liquid equilibria over the whole composition range confirm that reliable mixture properties can be obtained over a wide range of state points with simple molecular models when the unlike interactions are adjusted to a small experimental data basis.

\section{Conclusion}

A new rigid molecular model for ethanol was developed modelling dispersion and repulsion with three Lennard-Jones sites for the methyl, methylene and hydroxyl group and modelling the polarity and hydrogen bonding with three point charges located at the methylene and hydroxyl Lennard-Jones sites, as well as at the nucleus of the hydroxyl hydrogen. The parameters were optimized to give accurate pure component saturated liquid density and vapor pressure. This potential model yields mean unsigned errors in vapor pressure, saturated liquid density and heat of vaporization of 3.7, 0.3 and 0.9~\%, respectively, in the temperature range from 270 to 490 K. This new molecular model for ethanol was applied together with two-center Lennard-Jones plus pointquadrupole (2CLJQ) models \cite{Vrabec2001} for methane, nitrogen, oxygen and carbon dioxide to predict Henry's law constants. 

Using the Lorentz-Berthelot comining rule, simulations were performed to predict Henry's law constants of the four solutes in ethanol from 273 to 498~K with an increment of 25~K. For methane, nitrogen and oxygen, favorable predictions are achieved with deviations below 20~\%. For carbon dioxide, for which the Henry's law constants are small, deviations of up to 70~\% are observed.

As an alternative to predict from pure component data alone, the modified Lorentz-Berthelot combining rule was studied. The binary interaction parameter $\xi$ was adjusted to one experimental Henry's law constant at 298~K, a temperature for which reliable data on the Henry's law constants in ethanol are available for all studied solutes. The models with the adjusted binary interaction parameter give  what we believe is the presently most reliable information on $H(T)$ in the studied systems. The adequacy of the mixture models are supported by the vapor-liquid equilibrium simulation results for the mixture carbon dioxide and ethanol over the whole composition range. 

The methods applied in this work to calculate Henry's law constants are limited to solute molecules of smaller size than the solvent molecules due to Widom's test particle method. In future work, the feasibility with strongly dipolar solute molecules will be examined.   

\section{Acknowledgment}

The authors gratefully acknowledge financial support by Deutsche Forschungsgemeinschaft, Son\-der\-for\-schungs\-be\-reich 412, University of Stuttgart as well as Schwerpunktprogramm 1155 and the use of computational resources provided by High Performance Computing Center Stuttgart in the frame of the project MMHBF. Also we want to thank Bernhard Eckl for setting up the quantum chemical calculations.

\section{Appendix, Simulation Details \label{simdet}}

In all simulations $N=864$ molecules were used. The calculation of the vapor-liquid equilibra of pure ethanol was done with the $NpT$+test particle method \cite{Moeller1990}-\cite{Moeller1994}. In order to determine the chemical potential and pure ethanol properties in the liquid phase, the gradual insertion method \cite{Shevkunov1988}-\cite{Vrabec2002a} was applied. In such a simulation run, the liquid is equilibrated in the $NpT$ ensemble sufficiently over 10~000 cycles without fluctuating molecules and adjusting the maximum displacement of  translation, rotation and volume to yield acceptance rates of 50~\%. After that, 5~000 Monte Carlo cycles with fluctuating molecules are performed to equilibrate the weights of the transition states. The production phase was performed over 50~000 Monte Carlo cycles. The nonentropic properties and the chemical potential in the vapor phase were determined applying standard Monte Carlo simulation and Widom's test particle insertion \cite{Widom1963}. A vapor simulation consists of a sufficiently long equilibration phase and 200~000 Monte Carlo cycles where $N$ test particles are inserted after each cycle.

In order to evaluate the Henry's law constants, molecular dynamics simulations have been performed. After a long equilibration of 200~000 time steps, 250~000 production time steps were carried out inserting $4N$ test particles after each step.

The Lennard-Jones long range interactions beyond the cut-off radius were corrected employing angle averaging proposed by Lustig \cite{Lustig1988}. The coulombic interactions were corrected using the reaction field method \cite{Allen1987}. The cutoff radius was at least 17.5 \r{A}. The quadrupolar interaction needs no long range correction, as its contribution disappears by orientational averaging. For the simulation of pure ethanol vapor, the diameter of the cutoff sphere was chosen close to half simulation box length due to faster equilibration and better sampling.

\clearpage



\clearpage

\listoffigures
\clearpage



\begin{table}[ht]
\noindent
\caption[]{Lennard-Jones, point charge and geometry parameters of the present ethanol model, cf.~Equation (\ref{equ_etha}) and Figure \ref{geo_etha}; electronic charge $e=1.602177 \cdot 10^{-19}$ C.\label{tab_etha_pot}}
\bigskip
\begin{center}
\begin{tabular}{|l|c|c|c|} \hline
Site                      & $\sigma_{aa}$ & $\epsilon_{aa}/k_{\rm B}$ & $q_{ia}$    \\
                          & \r{A}         & K                       & $ e $  \\ \hline
$\rm{S}_{\rm{CH3}}$   & 3.6072      & 120.15                &    0       \\
$\rm{S}_{\rm{CH2}}$   & 3.4612      & 86.291                & \multicolumn{1}{|r|}{   0.25560} \\
$\rm{S}_{\rm{OH}} $   & 3.1496      & 85.053                & \multicolumn{1}{|r|}{$-$0.69711}  \\
$\rm{S}_{\rm{H}}  $   & 0           &   0                   & \multicolumn{1}{|r|}{   0.44151}   \\ \hline
\end{tabular}

\vspace{4ex}

\begin{tabular}{|c|c|c|c|c|} \hline
$h_{1}$  & $h_{2}$  & $h_{3}$  & $\gamma_{1}$ & $\gamma_{2}$  \\
\r{A}    & \r{A}    &  \r{A}   & \mbox{deg}   &  \mbox{deg}    \\ \hline
1.98420  & 1.71581  & 0.95053 &  90.950  & 106.368          \\ \hline
\end{tabular}
\end{center}
\end{table}

\begin{table}[ht]
\noindent
\caption[]{Vapor-liquid equilibrium of pure ethanol: simulation results compared to experimental data \cite{Daubert1984} for vapor pressure, saturated densities and heat of vaporization. The numbers in parenthesis indicate the statistical uncertainty in last digit.}
\label{tab_eqr_etha}
\bigskip
\begin{center}
\begin{tabular}{|c||c|c|c|c|c|c|c|c|} \hline
&&&&&&&&\\[-2.0ex]
 $T$  &  $p_{\rm{sim}}$ & $p_{\rm{exp}}$ & $\rho'_{\rm{sim}}$ & $\rho'_{\rm{exp}}$ & $\rho''_{\rm{sim}}$ & 
 $\rho''_{\rm{exp}}$ & $\Delta h^{\rm{v}}_{\rm{sim}}$ & $\Delta h^{\rm{v}}_{\rm{exp}}$ \\[0.5ex]
K  & MPa & MPa & mol/l & mol/l &  mol/l & mol/l & kJ/mol & kJ/mol \\ \hline

270 & 0.00125(4)    & 0.00129 & 17.68(1) & 17.64 & 0.0008(4)    & 0.00056 & 45.03(4)  & 44.78 \\
314 & 0.01951(7)    & 0.01868 & 16.83(1) & 16.76 & 0.0077(4)    & 0.00710 & 42.03(4)  & 42.33 \\
358 & 0.1239\0(5)   & 0.12977 & 15.84(1) & 15.78 & 0.044\0(1)   & 0.04514 & 38.32(4)  & 38.80 \\
402 & 0.526\0\0(3)  & 0.55451 & 14.66(1) & 14.65 & 0.180\0(1)   & 0.18623 & 33.40(6)  & 33.99 \\
446 & 1.60\0\0\0(3) & 1.69564 & 13.16(2) & 13.24 & 0.56\0\0(1)  & 0.58896 & 27.1\0(1) & 27.48 \\
490 & 3.88\0\0\0(8) & 4.08339 & 11.10(4) & 11.14 & 1.5\0\0\0(2) & 1.68190 & 19.\0\0(1) & 17.96 \\ \hline

\end{tabular}
\end{center}
\end{table}

\clearpage

\begin{table}[ht]
\noindent
\caption[]{Parameter set of the 2CLJQ potential for the solutes, taken from \cite{Vrabec2001,Vrabec2004}.}
\label{tab_2cljq}
\bigskip
\begin{center}
\begin{tabular}{|l||c|c|c|c|} \hline
Solute & $\sigma$ & $\epsilon/k_{\rm B}$ & $L$   & $Q$   \\
      & \r{A}    & K                    & \r{A} & D\r{A} \\ \hline
$\rm CH_4$    & 3.7281 & 148.55 & --     & 0         \\ 
$\rm N_2$     & 3.3211 & 34.897 & 1.0464 & $-$1.4397 \\
$\rm O_2$     & 3.1062 & 43.183 & 0.9699 & $-$0.8081 \\ 
$\rm CO_2$    & 2.9847 & 133.22 & 2.4176 & $-$3.7938 \\ \hline
\end{tabular}
\end{center}
\end{table}

\begin{table}[ht]
\noindent
\caption[]{Henry's law constants of four solutes in ethanol for the predictive mode ($\xi=1$). The numbers in parenthesis indicate the statistical uncertainty in the last digits.}
\label{tab_pred}
\bigskip
\begin{center}
\begin{tabular}{|c||l@{ }r|l@{ }r|l@{ }r|l@{ }r|} \hline
$T$ & \multicolumn{2}{|c|}{$H_{\rm CH4}$} & \multicolumn{2}{|c|}{$H_{\rm N2}$} & \multicolumn{2}{|c|}{$ H_{\rm O2}$} & \multicolumn{2}{|c|}{$ H_{\rm CO2}$} \\
 K  & \multicolumn{2}{|c|}{bar}& \multicolumn{2}{|c|}{bar}  &    \multicolumn{2}{|c|}{bar} & \multicolumn{2}{|c|}{bar}  \\ \hline
273 & \0923&(16)   & 3647&(56)   & 1513&(19)     & 221.1&(38) \\
298 & \0997&(14)   & 3381&(43)   & 1542&(15)     & 302.4&(39) \\  
323 & 1076&(12)    & 3177&(32)   & 1593&(12)     & 390.2&(39) \\
348 & 1069.3&(79)  & 2810&(20)   & 1540.0&(84)   & 456.5&(30) \\
373 & 1070.9&(67)  & 2536&(16)   & 1491.8&(73)   & 515.3&(29) \\
398 & 1007.8&(57)  & 2171&(12)   & 1374.7&(64)   & 540.7&(28) \\
423 & \0929.9&(58) & 1835&(12)   & 1236.9&(73)   & 547.2&(31) \\
448 & \0817.8&(39) & 1483.0&(78) & 1068.1&(46)   & 524.1&(23) \\
473 & \0703.2&(44) & 1172.4&(84) & \0896.3&(54)  & 485.5&(27) \\
498 & \0552.1&(43) & \0837.4&(77)& \0688.4&(55)  & 411.9&(28)  \\ \hline
\end{tabular}
\end{center}
\end{table}

\begin{table}[ht]
\noindent
\caption[]{Henry's law constants of four solutes in ethanol for the adjusted mode: methane $\xi$=1.0403, nitrogen $\xi$=1.0449, oxygen $\xi$=0.9802 and carbon dioxide $\xi$=1.0790. The numbers in parenthesis indicate the statistical uncertainty in the last digits.}
\label{tab_desc}
\bigskip
\begin{center}
\begin{tabular}{|c||l@{ }r|l@{ }r|l@{ }r|l@{ }r|} \hline
$T$ & \multicolumn{2}{|c|}{$H_{\rm CH4}$} & \multicolumn{2}{|c|}{$H_{\rm N2}$} & \multicolumn{2}{|c|}{$ H_{\rm O2}$} & \multicolumn{2}{|c|}{$ H_{\rm CO2}$} \\
 K  & \multicolumn{2}{|c|}{bar}& \multicolumn{2}{|c|}{bar}  &    \multicolumn{2}{|c|}{bar} & \multicolumn{2}{|c|}{bar}  \\ \hline
273 & 686&(12)   & 2867&(43)   & 1691&(17)   & 101.7&(17) \\ 
298 & 797&(11)   & 2827&(34)   & 1765&(16)   & 158.1&(20) \\ 
323 & 848.4&(89) & 2614&(26)   & 1763&(14)   & 213.8&(20) \\ 
348 & 877.1&(82) & 2390&(21)   & 1691&(11)   & 269.1&(22) \\ 
373 & 878.0&(59) & 2148&(14)   & 1601.0&(99) & 319.5&(19) \\ 
398 & 865.0&(51) & 1942&(11)   & 1483.4&(69) & 360.3&(19) \\ 
423 & 809.4&(43) & 1636&(11)   & 1306.1&(70) & 381.2&(20) \\ 
448 & 733.7&(38) & 1343.2&(69) & 1117.9&(49) & 383.6&(15) \\ 
473 & 634.0&(38) & 1078.2&(75) & \0931.6&(57)& 372.7&(20) \\ 
498 & 507.9&(38) & \0783.3&(70)& \0705.0&(57)& 332.8&(20) \\ \hline
\end{tabular}
\end{center}
\end{table}
\clearpage


\clearpage

\clearpage

\clearpage

\clearpage

\clearpage

\clearpage

\clearpage

\clearpage

\end{document}